\documentclass[RNAAS]{aastex62}

\begin{document}

\title{An edge-on disk in the quadruply lensed quasar cross GraL~J181730853+272940139}

\correspondingauthor{Cristian E. Rusu}
\email{cerusu@naoj.org}

\author[0000-0003-4561-4017]{Cristian E. Rusu}
\altaffiliation{Subaru Fellow}
\affiliation{Subaru Telescope, National Astronomical Observatory of Japan, 650 N Aohoku Pl, Hilo, HI 96720}


\author{Cameron A. Lemon}
\altaffiliation{}
\affiliation{Institute of Astronomy, University of Cambridge, Madingley Road, Cambridge CB3 0HA, UK}
\affiliation{Kavli Institute for Cosmology, University of Cambridge, Madingley Road, Cambridge CB3 0HA, UK}

\keywords{gravitational lensing: strong: individual: GraL~J181730853+272940139}

\section{Introduction} 

Data mining through large, multi-band ground-based surveys and the recent availability of source catalogues from the {\it Gaia} space mission \citep[e.g.,][]{gaia18} has led to a surge in the number of known gravitationally lensed quasars \citep[e.g.,][]{lemon18,delchambre18}, including some unusual systems, such as with a central 5th image \citep{ostrovski18}, exceedingly strong flux anomaly \citep{glikman18}, or exceedingly large external shear \citep{rusu18}.

One of these newly discovered systems is GraL~J181730853+272940139, hereafter J1817+2729. This system was reported in \citet{delchambre18}, and was identified as a quadruply lensed quasar candidate via a machine learning search inside {\it Gaia} Data Release 2, where it consists of three sources with maximum separation $1.8\arcsec$. Lemon et al. in prep (private communications) confirmed it spectroscopically as a lensed quasar with source redshift $z=3.07$. A mass model of this system has not been published to date, and the highest resolution imaging data available is from PanSTARRS1 \citep{chambers16}, with seeing $\sim1.0\arcsec-1.3\arcsec$ in the $grizy$ filters.

Here we present the first mass model of J1817+2729, which includes the measured relative astrometry and morphology of the lensing galaxy, using newly acquired Subaru/FOCAS \citep{kashikawa02} imaging. 

\section{Imaging data and mass modeling} 

We acquired $4\times180$s frames in $I-$band with sub-arcsecond seeing and pixel scale $0.1038\arcsec$, on 2018 August 18. We performed standard reduction with \texttt{FOCASRED} \footnote{\url{https://subarutelescope.org/Observing/Instruments/FOCAS/Detail/UsersGuide/DataReduction/focasred.html}}. The best frame has a seeing of $0.84\arcsec$, and we used it to measure the photometry for all components of J1817+2729, as well as the morphology of the lensing galaxy. Accurate relative astrometry is crucial for obtaining a robust mass model, but is usually unreliable when measured from a single image with relatively large seeing \citep[e.g.,][]{rusu16}. We therefore took into account the scatter in the measured astrometry from the two best frames. We calibrated the photometry to the Vega system by measuring the magnitude of a star nearest J1817+2729 in the FOCAS data as well as archival PanSTARRS1 images, and employing color transformations from \citet{waters16}. We measured astrometry/photometry/morphology with \texttt{hostlens} \citep{rusu16}, using nearby stars as point-spread functions and a De Vaucouleurs profile for the lensing galaxy, to ensure convergence. We fitted the lensing configuration with \texttt{glafic} \citep{oguri10}, at first by using a singular isothermal ellipsoid profile with external shear, constrained by the relative astrometry of the lens and images, and the flux ratio of images A/B (assuming 20\% uncertainty). The redshift of the lensing galaxy is unknown and difficult to estimate photometrically, as the lens is only detected in the FOCAS $I-$band  and the PanSTARRS1 $i-$band.

\begin{figure}[]
\begin{center}
\includegraphics[width=\textwidth,angle=0]{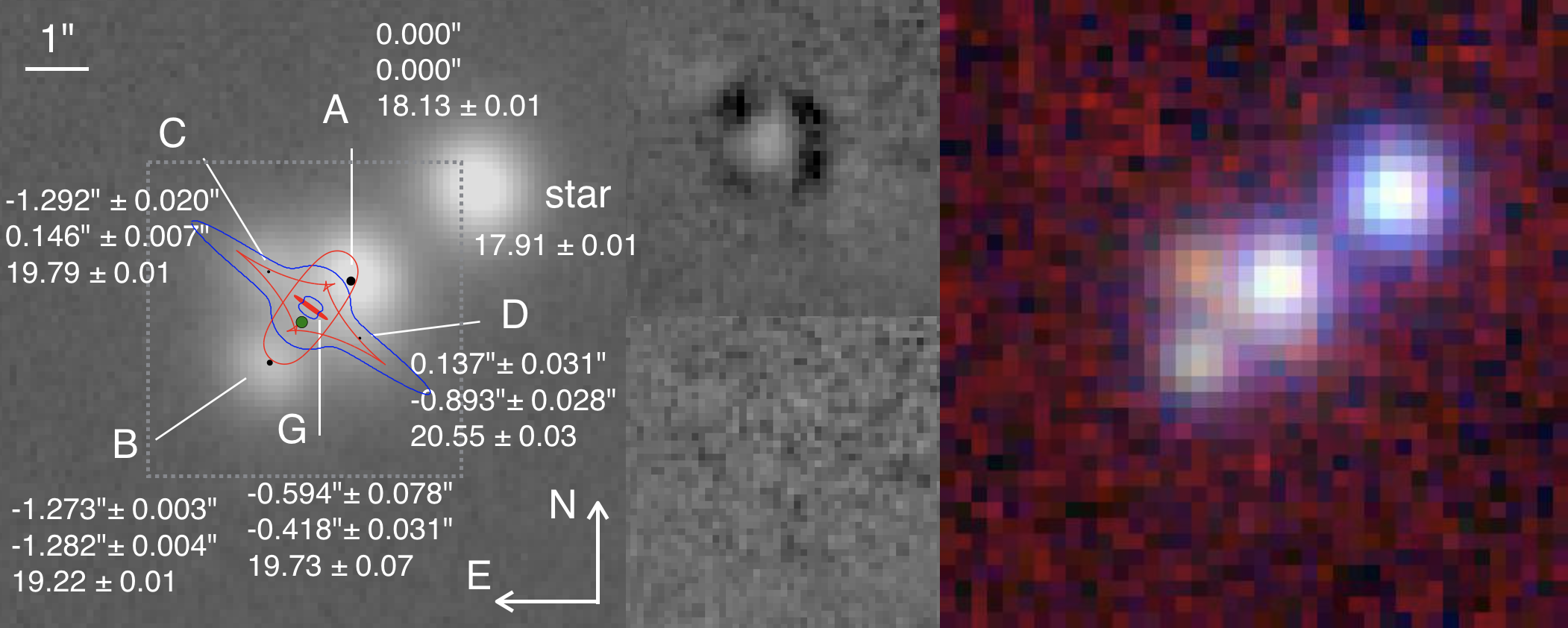}
\caption{{\it Left:} $10\arcsec\times10\arcsec$ Subaru/FOCAS $I-$band image of J1817+2729. The three rows of text next to each component show the measured relative astrometry along the E-W axis, the S-N axis, and the absolute photometry (not including zero-point uncertainties), respectively. The blue and red contours mark the critical lines and caustics of the mass model. The black circles mark the positions of the quasar images predicted by the model, with radius scaled in proportion to the predicted fluxes. The red ellipsis marks the measured position of the lensing galaxy, with axis ratio given by the mass model. The green circle marks the position of the source quasar. The dotted square marks the central $5\arcsec\times5\arcsec$ region, for which the residuals after subtracting the best-fit light model without ({\it upper}) and accounting for ({\it lower}) a single De Vaucouleurs component lensing galaxy are shown in the central panel, on the same logarithmic scale. {\it Right:} color combined image of the system from Pan-STARRS1. \label{fig}}
\end{center}
\end{figure}

We show the FOCAS image of J1817+2729 in Figure \ref{fig}, left panel. Our \texttt{hostlens} modeling clearly detects the lensing galaxy, which would otherwise appear as outstanding residuals (upper center panel). The lensing galaxy has a large ellipticity of $0.67\pm 0.02$, a major axis oriented at $54\pm1$ deg E of N, and an effective radius along the major axis of $0.71\arcsec\pm0.03\arcsec$. Our mass modeling produces a statistically good fit with reduced $\chi^2=1.6$ for 2 degrees of freedom (dof). However, the model is unusual, as it requires a very large shear $\gamma\sim0.28$, orthogonal to the major axis of the lensing galaxy, and high ellipticity $e\sim0.83$. This suggests that the lens is more complex, e.g. with a dominant edge-on disk. Indeed, remodeling the lensing galaxy light with a concentric exponential disk and a De Vaucouleurs bulge reveals an edge-on disk with $e\sim0.92$, and effective radius $\sim0.3\arcsec$. We adopt a mass model consisting of concentric exponential profile (for the edge-on disk) and a singular isothermal sphere with external shear (for the bulge + dark matter halo), where the only morphological constraint we impose is the robust orientation of the disk. This produces a reduced $\chi^2=0.3$ for 1 dof, a ratio of the Einstein radii of the two components $\sim2.3$, $e=0.85\pm0.07$, effective radius $0.46\arcsec\pm0.08\arcsec$, and a much reduced shear of $0.04\pm0.02$ at $-18.4\pm4.4$ deg.
The critical lines and caustics for this model are overlaid on Figure \ref{fig}. Finally, the PanSTARRS1 imaging (Figure \ref{fig}, right) reveals that images C and D are reddened. This is consistent with the orientation of the lensing galaxy, which goes across them, and suggests a dusty disk. We conclude that the large quadrupole in this system is caused by the dominant effect of the mass in the disk. 

\acknowledgments
We thank Paul Schechter for making this work possible, and Jen-Wei Hsueh for useful comments. {\it Facilities and archival data:} Subaru/FOCAS, PanSTARRS1.


\begin{thebibliography}{}

\bibitem[Chambers et al.(2016)]{chambers16} Chambers, K.~C., Magnier, E.~A., Metcalfe, N., et al.\ 2016, arXiv:1612.05560 
\bibitem[Delchambre et al.(2018)]{delchambre18} Delchambre, L., Krone-Martins, A., Wertz, O., et al.\ 2018, arXiv:1807.02845 
\bibitem[Gaia Collaboration et al.(2018)]{gaia18} Gaia Collaboration, Brown, A.~G.~A., Vallenari, A., et al.\ 2018, \aap, 616, A1 
\bibitem[Glikman et al.(2018)]{glikman18} Glikman, E., Rusu, C.~E., Djorgovski, S.~G., et al.\ 2018, arXiv:1807.05434 
\bibitem[Kashikawa et al.(2002)]{kashikawa02} Kashikawa, N., Aoki, K., Asai, R., et al.\ 2002, \pasj, 54, 819 
\bibitem[Lemon et al.(2018)]{lemon18} Lemon, C.~A., Auger, M.~W., McMahon, R.~G., \& Ostrovski, F.\ 2018, \mnras, 479, 5060 
\bibitem[Oguri(2010)]{oguri10} Oguri, M.\ 2010, \pasj, 62, 1017
\bibitem[Ostrovski et al.(2018)]{ostrovski18} Ostrovski, F., Lemon, C.~A., Auger, M.~W., et al.\ 2018, \mnras, 473, L116 
\bibitem[Rusu et al.(2016)]{rusu16} Rusu, C.~E., Oguri, M., Minowa, Y., et al.\ 2016, \mnras, 458, 2 
\bibitem[Rusu et al.(2018)]{rusu18} Rusu, C.~E., Berghea, C.~T., Fassnacht, C.~D., et al.\ 2018, arXiv:1803.07175 
\bibitem[Waters et al.(2016)]{waters16} Waters, C.~Z., Magnier, E.~A., Price, P.~A., et al.\ 2016, arXiv:1612.05245

\end{thebibliography}
\end{document}